\documentclass[aps,reprint,showkeys,showpacs]{revtex4-1}
\usepackage{amssymb,amsfonts}
\usepackage{amsmath,bbm}
\usepackage{fourier}
\usepackage[T1]{fontenc}
\input xy
\xyoption{all}
\usepackage[pdftex]{graphicx}

\newcommand{\CC}{\mathbb C}

\newcommand{\U}{\mathcal{U}}
\renewcommand{\u}{\mathfrak{u}}
\newcommand{\I}{\mathbbm{I}}
\renewcommand{\1}{{\mathbf 1}}
\newcommand{\0}{{\mathbf 0}}

\newcommand{\tr}{\operatorname{Tr}}
\newcommand{\bdist}{\operatorname{dist_{B}}}
\newcommand{\Dinv}{\mathcal{D}_\text{inv}}
\newcommand{\Sinv}{\mathcal{S}_\text{inv}}
\newcommand{\diag}{\operatorname{diag}}
\newcommand{\Ker}{\operatorname{Ker}}
\newcommand{\T}{\operatorname{T}\!}
\renewcommand{\H}{\operatorname{H}\!}
\newcommand{\V}{\operatorname{V}\!}

\newcommand{\HH}{\mathcal{H}}
\newcommand{\D}{\mathcal{D}}
\newcommand{\A}{\mathcal{A}}
\renewcommand{\L}{\mathcal{L}}
\renewcommand{\S}{\mathcal{S}}

\newcommand{\half}{\frac 12}

\newcommand{\dt}{\operatorname{d}\!t}

\newcommand{\eps}{\varepsilon}

\newcommand{\dd}[1]{\frac{\operatorname{d}}{\operatorname{d}\!#1}}
\newcommand{\length}[1]{\operatorname{Length}[#1]}

\newcommand{\dist}[2]{\operatorname{dist}(#1,#2)}

\begin{document}
\title{Geometry of quantum dynamics and a time-energy uncertainty relation for mixed states}
\date{\today}
\author{Ole Andersson}
\email{olehandersson@gmail.com}
\affiliation{Department of Physics, Stockholm University, 10691 Stockholm, Sweden}
\author{Hoshang Heydari}
\email{Hoshang@fysik.su.se}
\affiliation{Department of Physics, Stockholm University, 10691 Stockholm, Sweden}

\begin{abstract}
In this paper we establish important relations between Hamiltonian
dynamics and Riemannian structures on phase spaces for unitarily evolving
finite level quantum systems in mixed states. We show that the 
energy dispersion (i.e. $1/\hbar$ times the path integral of the 
energy uncertainty) of a unitary evolution is bounded from below by 
the length of the evolution curve. Also, we show that for each curve of mixed states 
there is a Hamiltonian for which the curve is a solution to the 
corresponding von Neumann equation, and the energy dispersion equals 
the curve's length. This allows us to express the distance between two
mixed states in terms of a measurable quantity, and derive a time-energy 
uncertainty relation for mixed states. In a final section we compare our results with an 
energy dispersion estimate by Uhlmann. 
\end{abstract}

\maketitle

\section{Introduction}
Ever since the advent of general relativity, scientists have been looking for geometrical principles underlying physical laws.
Nowadays it is well known that geometry 
affects the physics on all length scales, and physical theory building consists to a large extent of geometrical considerations.
This paper concerns geometric quantum mechanics, a branch of quantum physics that has received much attention lately (which is largely due to the crucial role geometry plays in quantum information and quantum computing \cite{Pachos_etal1990, *Zanardi_etal1999, *Ekert_etal2000, *Zanardi_etal2007, *Rezakhani_etal2010, Jones_etal2000, *Falci_etal2000, *Farhi_etal2001, *Duan_etal2001, *Recati_etal2002}).
Here we equip the phase spaces
for unitarily evolving finite level quantum systems with natural Riemannian structures, and 
establish remarkable but fundamental relations between these and Hamiltonian dynamics.

A quantum system prepared in a pure state is usually modeled on a projective Hilbert space, and if the system is closed its state will evolve unitarily in this space.
Aharonov and Anandan \cite{Anandan_etal1990} showed that for unitary evolutions there is a geometric quantity which, like Berry's celebrated phase \cite{Berry1984,Simon1983}, is independent of the particular Hamiltonian used to transport a pure state along a given route. 
More precisely, they showed that
the energy dispersion (i.e. $1/\hbar$ times the path integral of the energy uncertainty) of an evolving state equals the Fubini-Study length of
the curve traced out by the state. 
Using this, Aharonov and Anandan gave a new geometric interpretation of the time-energy uncertainty relation. 

The state of an experimentally prepared quantum system  generally
exhibits classical uncertainty, and is most appropriately described as a probabilistic mixture of pure states. It is common to represent mixed states by density operators, and many metrics on spaces of density operators have been developed to capture various physical, mathematical, or information theoretical aspects of quantum mechanics \cite{Bengtsson_etal2008,Nielsen_etal2010}.
In this paper we utilize a construction by Montgomery \cite{Montgomery1991} to provide the spaces of isospectral density operators with Riemannian metrics, and we show that these metrics admit a generalization of the energy dispersion result of Aharonov and Anandan to evolutions of finite dimensional quantum systems in mixed states. 
Indeed, we show that the energy dispersion of an evolving mixed state is bounded from below by the length of the curve traced out by the density operator of the state, and we show that 
every curve of isospectral density operators is generated by a Hamiltonian 
for which the energy dispersion equals the curve's length. The latter result allows us to express the distance between two mixed states in terms of a measurable quantity, and we use it to derive a time-energy uncertainty principle for mixed states.

Uhlmann \cite{Uhlmann1986, Uhlmann1991} was among the first to develop a mathematical framework similar to the one presented here. 
In \cite{Uhlmann1992Energy} he used it to derive an estimate for the energy dispersion of an evolving mixed state. We compare our energy dispersion estimate with that of Uhlmann in this paper's final section.

\section{Geometry of orbits of isospectral density operators}
In this paper we will only be interested in  finite dimensional quantum systems that evolve unitarily. 
They will be modeled on a Hilbert space $\HH$ of unspecified dimension $n$, and their states will be represented by density operators.
Recall that a density operator is a Hermitian, nonnegative operator with unit trace.
We write $\D(\HH)$ for the space of density operators on $\HH$.

\subsection{Riemannian structure on orbits of density operators}
A density operator whose evolution is governed by a von Neumann equation remains in a single orbit of the left conjugation action of the unitary group of $\HH$ on $\D(\HH)$. The orbits of this action are in one-to-one correspondence with the possible spectra for density operators on $\HH$, where by the \emph{spectrum} of a density operator of rank $k$ we mean the decreasing sequence
\begin{equation}
\sigma=(p_1,p_2,\dots,p_k)
\label{spectrum}
\end{equation}
of its, not necessarily distinct, positive eigenvalues.
Throughout this paper we fix $\sigma$, and write $\D(\sigma)$ for the corresponding orbit.

To furnish $\D(\sigma)$ with a geometry, let $\L(\CC^k,\HH)$ be the space of linear maps from $\CC^k$ to $\HH$ 
equipped with the Hilbert-Schmidt Hermitian inner product, and $P(\sigma)$ be 
the diagonal $k\times k$ matrix 
that has $\sigma$ as its diagonal.
Inspired by Montgomery \cite{Montgomery1991}, we set
\begin{equation*}
\S(\sigma)=\{\Psi\in\L(\CC^k,\HH):\Psi^\dagger \Psi=P(\sigma)\},
\end{equation*}
and define  
\begin{equation*}
\pi:\S(\sigma)\to\D(\sigma),\quad \Psi\mapsto\Psi\Psi^\dagger.
\label{bundle}
\end{equation*}
Then $\pi$ is a principal fiber bundle with right acting gauge group
\begin{equation*}
\U(\sigma)
=\{U\in\U(k):UP(\sigma)=P(\sigma)U\},
\end{equation*}
whose Lie algebra is
\begin{equation*}
\u(\sigma)
=\{\xi\in\u(k):\xi P(\sigma)=P(\sigma)\xi\}.
\end{equation*}
The real part of the Hilbert-Schmidt product restricts to a gauge invariant Riemannian metric $G$ on $\S(\sigma)$,
\begin{equation*}
G(X,Y)=\half\tr(X^\dagger Y+Y^\dagger X),
\end{equation*}
and we equip $\D(\sigma)$ with the unique metric $g$ that makes $\pi$ a Riemannian submersion.

\subsection{Mechanical connection}
The \emph{vertical} and \emph{horizontal bundles} over $\S(\sigma)$ are the subbundles 
$\V\S(\sigma)=\Ker \pi_*$ and $\H\S(\sigma)=\V\S(\sigma)^\bot$ 
of the tangent bundle of $\S(\sigma)$. Here $\pi_*$ is the differential of $\pi$ and $^\bot$ denotes orthogonal complement with respect to $G$.
Vectors in $\V\S(\sigma)$ and $\H\S(\sigma)$
are called vertical and horizontal, respectively,
and a curve in $\S(\sigma)$ is called horizontal if its velocity vectors are horizontal.
Recall that for every curve $\rho$ in $\D(\sigma)$ and every $\Psi_0$ in the fiber over $\rho(0)$ there is a unique horizontal lift of $\rho$ to $\S(\sigma)$ that extends from $\Psi_0$. 
This lift and $\rho$ have the same lengths, since $\pi$ is a Riemannian submersion.

The infinitesimal generators of the gauge group action yield canonical isomorphisms between $\u(\sigma)$ and the fibers in $\V\S(\sigma)$:
\begin{equation}\label{eq:inf gen}
\u(\sigma)\ni\xi\mapsto \Psi\xi\in\V_\Psi\S(\sigma).
\end{equation}
Furthermore, $\H\S(\sigma)$ is the kernel bundle of the gauge invariant \emph{mechanical connection form}
$\A_{\Psi}=\I_{\Psi}^{-1}J_{\Psi}$,
where $\I_{\Psi}:\u(\sigma)\to \u(\sigma)^*$ and $J_{\Psi}:\T_{\Psi}{\S(\sigma)}\to \u(\sigma)^*$ are the \emph{moment of inertia} and \emph{moment map}, respectively,
\begin{equation*}
\I_{\Psi}\xi\cdot \eta=G(\Psi\xi,\Psi\eta),\quad 
J_{\Psi}(X)\cdot\xi=G(X,\Psi\xi).
\end{equation*}
The moment of inertia is of \emph{constant bi-invariant type} since it is an adjoint-invariant form on $\u(\sigma)$ which is independent of $\Psi$ in $\S(\sigma)$. To be exact,
\begin{equation}\label{eq:beta}
\I_{\Psi}\xi\cdot \eta=\half\tr\left(\left(\xi^\dagger \eta+\eta^\dagger \xi\right)P(\sigma)\right).
\end{equation}
Using \eqref{eq:beta} we can derive an explicit formula for the connection form.
Indeed, if $m_1, m_2, \dots , m_l$ are the multiplicities of the different eigenvalues in $\sigma$, with $m_1$ being the multiplicity of the greatest eigenvalue, $m_2$ the multiplicity of the second greatest eigenvalue, etc., and if for $j=1,2,\dots,l$, 
\begin{equation*}
E_j=\diag(\0_{m_1},\dots,\0_{m_{j-1}},\1_{m_j},\0_{m_{j+1}},\dots,\0_{m_l}),
\end{equation*}
then
\begin{equation*}
\begin{split}
\I_\Psi\Big(&\sum_jE_j\Psi^\dagger XE_jP(\sigma)^{-1}\Big)\cdot\xi=\\
&=\half\tr\Big(\sum_jE_jX^\dagger\Psi E_j\xi-\xi E_j\Psi^\dagger XE_j\Big)\\
&=\half\tr\big(X^\dagger\Psi\xi-\xi\Psi^\dagger X\big)\\
&=J_\Psi(X)\cdot\xi
\end{split}
\end{equation*}
for every $X$ in $\T_\Psi\S(\sigma)$ and every $\xi$ in $\u(\sigma)$.
Hence
\begin{equation*}\label{eq:explicit}
\A_\Psi(X)=\sum_jE_j\Psi^\dagger XE_jP(\sigma)^{-1}.
\end{equation*}
Observe that the orthogonal projection of $\T_\Psi\S(\sigma)$ onto $\V_\Psi\S(\sigma)$ 
is given by the connection form followed by the infinitesimal generator \eqref{eq:inf gen}. Therefore, the \emph{vertical} and \emph{horizontal projections} of $X$ in $\T_\Psi\S(\sigma)$ are $X^\bot=\Psi\A_\Psi(X)$ and $X^{||}=X-\Psi\A_\Psi(X)$, respectively.


\section{Geometrical uncertainty estimates}

If $\hat A$ is an observable on $\HH$, 
the gauge invariant vector field $X_{\hat A}$ on $\S(\sigma)$ is defined by
\begin{equation*}
X_{\hat A}(\Psi)=\dd\eps\left[\exp\left(\frac{\eps}{i\hbar}\hat A\right)\Psi\right]_{\eps=0}.
\end{equation*}
Let $X_A$ be the projection of  $X_{\hat A}$ onto $\D(\sigma)$, and define the \emph{uncertainty of $\hat A$} to be the scalar field $\Delta A$ on $\D(\sigma)$ given by 
\begin{equation*}
\Delta A(\rho)=\sqrt{\tr(\hat A^2\rho)-\tr(\hat A\rho)^2}.
\end{equation*}
We will show that
\begin{align}
&\Delta A(\rho)\geq \hbar \sqrt{g(X_A(\rho),X_A(\rho))},\label{main ett}\\
&\Delta A(\rho)=\hbar\sqrt{g(X_A(\rho),X_A(\rho))}\text{ if } X_{\hat{A}}(\Psi)\in\H_\Psi\S(\sigma),\label{main tva}
\end{align}
where $\Psi$ is any element in the fiber over $\rho$. 

Assertion \eqref{main tva} follows immediately from the observations 
\begin{align}
&\tr(\hat A^2\rho)=\hbar^2 G(X_{\hat A}(\Psi),X_{\hat A}(\Psi)),\label{alfa}\\
&\tr(\hat A\rho)=i\hbar\tr(\A(X_{\hat A}(\Psi))P(\sigma))\label{beta}.
\end{align}
 For if $X_{\hat{A}}(\Psi)$ is horizontal, then the right hand side of \eqref{alfa} equals $\hbar^2 g(X_{A}(\Psi),X_{A}(\Psi))$, and the right hand side of \eqref{beta} vanishes.
If, on the other hand, $X_{\hat A}(\Psi)$ is not horizontal, we must estimate the difference between 
$G(X^{\bot}_{\hat A}(\Psi),X^{\bot}_{\hat A}(\Psi))$ and $\tr(\hat{A}\rho)^2$. 
The identity
\begin{equation*}
G(X^{\bot}_{\hat A}(\Psi),X^{\bot}_{\hat A}(\Psi))
=-\tr(\A(X_{\hat A}(\Psi))^2P(\sigma)),\label{gamma}
\end{equation*}
together with \eqref{alfa} and \eqref{beta} yield
\begin{equation}
\begin{split}
\Delta A(\rho)^2
=\hbar^{2}&g(X_A(\rho),X_A(\rho))\\
&+\hbar^{2}\tr(\A(X_{\hat A}(\Psi))P(\sigma))^2\\
&-\hbar^{2}\tr(\A(X_{\hat A}(\Psi))^2P(\sigma)).
\end{split}
\label{eq:eq}
\end{equation}
Now \eqref{main ett} follows from the fact that the difference between the last two terms 
in \eqref{eq:eq} is nonnegative.
To see this let $U$ in $\U(\sigma)$ be such that 
$iU\A_\Psi(X_{\hat A}(\Psi))U^\dagger$ is a diagonal matrix, say
\begin{equation*}
iU\A_\Psi(X_{\hat A}(\Psi))U^\dagger=\diag(\lambda_1,\lambda_2,\dots,\lambda_k).
\end{equation*}
Then 
\begin{equation*}
\begin{split}
\tr(\A_\Psi(X_{\hat A}(\Psi))P(\sigma))^2
&= -\Big(\sum_jp_j \lambda_j\Big)^2\\
&\geq -\sum_jp_j\lambda_j^2\\
&=\tr(\A_\Psi(X_{\hat A}(\Psi))^2P(\sigma)),
\end{split}
\label{eq:eqn}
\end{equation*}
since $U$ commutes with $P(\sigma)$ and $x\mapsto x^2$ is convex.

\subsection{Distance and energy dispersion}
The distance  between two density operators with common spectrum $\sigma$ is defined to be the infimum of the lengths of all curves in $\D(\sigma)$ that connects them. We will show that for any two density operators $\rho_0$ and $\rho_1$ in $\D(\sigma)$, 
\begin{equation}
\dist{\rho_0}{\rho_1}=\frac{1}{\hbar}\inf_{\hat H}\int_{t_0}^{t_1}\!\Delta H(\rho)\dt,\label{avstand}
\end{equation}
where the infimum is taken over all Hamiltonians $\hat H$ for which the \emph{boundary value von Neumann equation} 
\begin{equation}
\dot\rho=X_H(\rho),\qquad \rho(t_0)=\rho_0, \qquad\rho(t_1)=\rho_1,\label{von Neumann}
\end{equation}
is solvable.

The length of a curve $\rho$ in $\D(\sigma)$, with domain $t_0\leq t\leq t_1$ is 
\begin{equation*}
\length{\rho}=
\int_{t_0}^{t_1}\!\sqrt{g(\dot\rho,\dot\rho)}\dt.\label{length}
\end{equation*}
If $ \dot\rho=X_H(\rho)$, for some Hamiltonian $\hat H$, then, by \eqref{main ett}, the length of $\rho$ is a lower bound for the   
\emph{energy dispersion}: 
\begin{equation}
\length{\rho}\leq\frac{1}{\hbar}\int_{t_0}^{t_1}\!\Delta H(\rho)\dt.
\label{enekvation}
\end{equation}
There is a Hamiltonian $\hat H$ that 
generates a horizontal lift of $\rho$, because the unitary group of $\HH$ acts transitively on $\L(\CC^k,\HH)$.
For such a Hamiltonian we have equality in \eqref{enekvation}. Moreover, we can take $\rho$ to be \emph{length minimizing}, in the sense that $\length{\rho}=\dist{\rho_0}{\rho_1}$, because $\D(\sigma)$ is compact and hence $g$ is complete. Then, 
\begin{equation}
\dist{\rho_0}{\rho_1}=\frac{1}{\hbar}\int_{t_0}^{t_1}\!\Delta H(\rho)\dt,
\label{finalen}
\end{equation}
by \eqref{main tva}. Assertion \eqref{avstand} follows from \eqref{enekvation} and \eqref{finalen}.

\subsection{Time-energy uncertainty relation}
Consider a quantum system with Hamiltonian $\hat{H}$, and suppose $\rho$ is a solution to \eqref{von Neumann}.
The \emph{time-average of $\Delta H$} is
\begin{equation*}
\langle \Delta H\rangle=\frac{1}{\Delta t}\int_{t_0}^{t_1}\Delta H(\rho)\dt,\quad \Delta t=t_1-t_0.
\end{equation*}
We will show that if $\rho_0$ and $\rho_1$ are distinguishable \cite{Englert1996,Markham_etal2008}, then
\begin{equation}
\langle \Delta H\rangle \Delta t\geq \frac{\pi\hbar}{2}.
\label{energytime}
\end{equation}
For density operators representing pure states, this reduces to the time-energy uncertainty relation in \cite{Anandan_etal1990}.

Let $\Psi_0$ in $\pi^{-1}(\rho_0)$ and $\Psi_1$ in $\pi^{-1}(\rho_1)$ be
such that $\dist{\rho_0}{\rho_1}=\dist{\Psi_0}{\Psi_1}$.
The operators $\rho_0$ and $\rho_1$ have orthogonal supports, being distinguishable, and the same is true for $\Psi_0$ and $\Psi_1$ 
since the the support of $\Psi_0$ equals the support of $\rho_0$, and likewise for $\Psi_1$ and $\rho_1$. A compact way to express this is
\begin{equation}
\Psi_0^\dagger\Psi_1=0,\quad \Psi_1^\dagger\Psi_0=0.
\label{vinkelrata}
\end{equation}

If we consider 
$\Psi_0$ and $\Psi_1$ elements in $\S(\CC^k,\HH)$, the unit sphere in $\L(\CC^k,\HH)$, they are a distance of $\pi/2$ apart. In fact, $\Psi(t)=\cos(t)\Psi_0+\sin(t)\Psi_1$, with domain $0\leq t\leq \pi/2$, is a length minimizing curve from $\Psi_0$ to $\Psi_1$ in $\S(\CC^k,\HH)$.
Consequently,  
\begin{equation}
\dist{\rho_0}{\rho_1}\geq \pi/2.\label{storre}
\end{equation} 
The uncertainty relation \eqref{energytime} follows from \eqref{enekvation} and  \eqref{storre}.
Also note that the estimate \eqref{storre} cannot be improved. Direct computations using \eqref{vinkelrata} yield $\Psi(t)^\dagger\Psi(t)=P(\sigma)$ and $\Psi(t)^\dagger\dot\Psi(t)=0$. Therefore, $\Psi(t)$ is a horizontal curve in $\S(\sigma)$, and hence \eqref{storre} is, in fact, an equality.

\section{Uhlmann's bundle and the Bures distance}\label{Uhlmann and Bures}
Uhlmann \cite{Uhlmann1992Energy} proved that for unitarily evolving quantum systems represented by invertible density operators, the 
energy dispersion is bounded from below by the Bures distance between the initial and final states.
This result, together with 
\eqref{avstand}, shows that on orbits of invertible density operators, the distance function associated with $g$ is bounded from below by the Bures distance.
Here we present an independent argument for this fact, and we give an example of two isospectral density operators between which the two metrics measure different distances.

Let $\Sinv(\CC^n,\HH)$ be the space of all 
invertible maps in $\L(\CC^n,\HH)$ with unit norm, and $\Dinv(\HH)$  be the space of all invertible density operators acting on $\HH$. Then $\Pi:\Sinv(\CC^n,\HH)\to\Dinv(\HH)$ defined by $\Pi(\Psi)=\Psi\Psi^\dagger$ is a $\U(n)$-bundle, which we call \emph{Uhlmann's bundle} since it first appeared in \cite{Uhlmann1991}.
The geometry of Uhlmann's bundle has been thoroughly investigated, and it is an important tool in quantum information theory, mainly due to its close relationship with the Bures metric \cite{Bures1969,Uhlmann1992}.

Uhlmann's bundle is equipped with the mechanical connection, which means that the horizontal bundle is the orthogonal complement of the vertical bundle with respect to the Hilbert-Schmidt metric. Moreover, the metric on $\Dinv(\HH)$ obtained by declaring $\Pi$ to be a Riemannian submersion, is the Bures metric \cite{Uhlmann1992}. We denote the associated distance function by $\bdist$.

Suppose $k=n$ in \eqref{spectrum}. Then $\S(\sigma)$ is a submanifold of $\Sinv(\CC^n,\HH)$. Moreover, the vertical bundle of $\S(\sigma)$ is subbundle of the restriction of the vertical bundle of $\Sinv(\CC^n,\HH)$ to $\S(\sigma)$.
However, no nonzero horizontal vector in Uhlmann's bundle is tangential to $\S(\sigma)$.
To see this, let $\Psi$ be any element in $\S(\sigma)$. Then $X$ in $\T_\Psi\Sinv(\CC^n,\HH)$ is horizontal, i.e. is annihilated by the mechanical connection of the Uhlmann bundle, if and only if \cite{Uhlmann1991}
\begin{equation}\label{Uhlmann parallel}
\Psi^\dagger X-X^\dagger\Psi=0.
\end{equation}
On the other hand, every $X$ in $\T_\Psi\S(\sigma)$ satisfies
\begin{equation}\label{relation}
\Psi^\dagger X+X^\dagger\Psi=0
\end{equation}
since $\Psi^\dagger\Psi=P(\sigma)$. Clearly, only the zero vector satisfies both \eqref{Uhlmann parallel} and \eqref{relation}.

The distance between $\rho_0$ and $\rho_1$ in $\D(\sigma)$ is never smaller than Bures distance between them. Indeed, every curve between $\pi^{-1}(\rho_0)$ and $\pi^{-1}(\rho_1)$ in $\S(\sigma)$ is a curve between $\Pi^{-1}(\rho_0)$ and $\Pi^{-1}(\rho_1)$ in $\Sinv(\CC^n,\HH)$, and since the metrics on the total spaces of the two bundles are induced from a common ambient metric we can conclude that 
\begin{equation}\label{bures inequality}
\dist{\rho_0}{\rho_1}\geq\bdist(\rho_0,\rho_1).
\end{equation}

Uhlmann \cite{Uhlmann1992} and Dittmann \cite{Dittmann1993, Dittmann1999} have derived explicit formulas for the Bures metric for density operators on finite dimensional Hilbert spaces.
For density operators on $\CC^2$ the formula reads
\begin{equation}\label{Dittmann}
\bdist(\rho,\rho+\delta\rho)^2=\frac 14\tr\big(\delta\rho\delta\rho+\frac {1}{\det\rho}(\delta\rho-\rho\delta\rho)^2\big).
\end{equation}
We use \eqref{Dittmann} to show that there are density operators $\rho_0$ and $\rho_1$ acting on $\CC^2$ for which 
the inequality in \eqref{bures inequality}
is strict.

Suppose $\sigma=(p_1,p_2)$, and let $\eps>0$. For $\eps$ small enough,
\begin{equation*}
\rho(t)=\begin{bmatrix} p_2\sin^2(\eps t)+p_1\cos^2(\eps t) & (p_2-p_1)\sin(\eps t)\cos(\eps t)\\ (p_2-p_1)\sin(\eps t)\cos (\eps t) & p_1\sin^2(\eps t)+p_2\cos^2(\eps t)
\end{bmatrix}
\end{equation*}
is a length minimizing curve in $\D(\sigma)$ between $\rho_0=\rho(0)$ and $\rho_1=\rho(1)$. Thus
\begin{equation*}
\dist{\rho_0}{\rho_1}=\length{\rho}=\eps. 
\end{equation*}
However, \eqref{Dittmann} yields
\begin{equation*}
\bdist(\rho_0,\rho_1)=\frac {p_1-p_2}{\sqrt{2}}|\sin\eps|\sqrt{2+\frac{(p_1-p_2)^2}{2p_1p_2}\sin^2\eps}.
\end{equation*}

\section{Conclusion}
In this paper we have utilized a construction due to Montgomery, to equip the 
spaces of isospectral density operators acting on a finite 
dimensional Hilbert space with Riemannian metrics, and we have established important relations 
between these and Hamiltonian quantum dynamics.
Indeed, we have proved that the
energy dispersion of a unitarily evolving density operator is bounded from below 
by the length of the curve traced out by the operator, and that
every curve of isospectral density operators can be generated by a
Hamiltonian such that the energy dispersion equals the curve's length. 
These facts allowed us to express the distance between two density operators in terms of 
a measurable physical quantity, and the paper culminated in a time-energy uncertainty estimate for mixed states. 
In a final section we have compared our energy dispersion results with an energy dispersion estimate by Uhlmann. 

We believe that our results have very interesting applications in 
optimal quantum control.
Such aspects of the theory developed here will be investigated by the authors in a forthcoming paper. 
There we will focus on the geometry of, and dynamics in, orbits of invertible density operators, and we will classify the Hamiltonians that drive states along evolution curves with minimal energy dispersion.

\acknowledgments{The second author acknowledge the financial support  by the Swedish Research Council (VR).}

\end{document}